# *In silico* Proteome Cleavage Reveals Iterative Digestion Strategy for High Sequence Coverage

Jesse G. Meyer

Department of Chemistry and Biochemistry

University of California, San Diego

9500 Gilman Drive

La Jolla, CA 92093-0378

jgmeyer@ucsd.edu

**Abstract**

In the post-genome era, biologists have sought to measure the complete complement of proteins, termed proteomics. Currently, the most effective method to measure the proteome is with shotgun, or bottom-up, proteomics, in which the proteome is digested into peptides that are identified followed by protein inference. Despite continuous improvements to all steps of the shotgun proteomics workflow, observed proteome coverage is often low; some proteins are identified by a single peptide sequence. Complete proteome sequence coverage would allow comprehensive characterization of RNA splicing variants and all post translational modifications, which would drastically improve the accuracy of biological models. There are many reasons for the sequence coverage deficit, but ultimately peptide length determines sequence observability. Peptides that are too short are lost because they match many protein sequences and their true origin is ambiguous. The maximum observable peptide length is determined by several analytical challenges. This paper explores computationally how peptide lengths produced from several common proteome digestion methods limit observable proteome coverage. Iterative proteome cleavage strategies are also explored. These simulations reveal that maximized proteome coverage can be achieved by use of an iterative digestion protocol involving multiple proteases and chemical cleavages that theoretically allow 91.1% proteome coverage.

## 1. Introduction

In the post genome era, biologists have sought system-wide measurements of RNA, proteins, and metabolites, termed transcriptomics, proteomics, and metabolomics, respectively. Shotgun, or bottom-up, proteomics has become the most comprehensive method for proteome identification and quantification [1]. However, observed protein sequence coverage is often low. The ability to cover 100% of protein sequences in a biological system was likened to surrealism in a recent review by Karas *et al* [2]. Multiple steps in the traditional shotgun proteomics workflow contribute to the deficit in observed sequence coverage, including: proteome isolation, proteome digestion, peptide separation, peptide MS/MS, and identification by peptide-spectrum matching. Proteome isolation has been extensively evaluated [3,4]. Several types of peptide separation have been have been explored [5,6,7]. Mass spectrometers are becoming more sensitive and versatile [8,9,10]. Peptide-spectrum matching algorithms are adapting to new data types [11] and becoming more sensitive [12,13]. Proteome fragmentation into sequence-able peptides is one step with significant room for improvement. DNA sequencing relies on sequence fragmentation into readable pieces by mechanical force [14], which produces a nearly uniform distribution of fragment lengths. In comparison, proteome fragmentation is generally accomplished by targeting one or more amino acid residues for cleavage, and therefore, the protein cleavage can be likened to a Poisson process that produces an exponential distribution of peptide lengths.

Numerous papers have described the application of new digestion strategies for proteome analysis [15,16,17,18], however, no single strategy has emerged as optimal.

The greatest observed proteome coverage has plateaued around 25%. 24.6% of the human proteome was recently observed [19], but this was obtained from over 1,000 MS/MS data files that allowed identification of over 260,000 peptide sequences using a new high performance data analysis package. Similarly, multiple protease digests of yeast resulted in 25.2% coverage [20]. Therefore, improved strategies for proteome digestion are needed to allow observation of a complete proteome.

An innovative example demonstrating the application of multiple enzyme digestion (MED) was recently published by Wiśniewski and Mann [21], which demonstrated the utility of multi-enzyme digestion coupled to filter-aided sample preparation [22] (MED-FASP, figure 1). This work extends a previous work that described size exclusion to isolate long tryptic peptides for additional digestion [18]. Wiśniewski and Mann compared gains afforded by iterative digestion using various proteases (i.e. GluC, ArgC, LysC, or AspN) followed by trypsin. Their work concluded that iterative digestion with LysC followed by Trypsin allowed 31% more protein identifications and a 2-fold gain in observed phosphopeptides for a particular protein. Their work led me to optimize iterative digestion *in silico* with the hope of identifying a testable digestion strategy that can theoretically achieve complete proteome coverage.

2. **Methods**

The *S. cerevisiae* proteome file in FASTA format was downloaded from uniprot on June 20[th], 2012. Proteome digestion simulations were accomplished using scripts written in [R] [23]. Considered protease specificities include c-terminal of: R/K (trypsin), L (LeuC theoretical cleavage agent), E (GluC), and K (LysC). Additionally, simulations utilized chemical digestion agents [24], including cyanogen bromide (CNBr) [25,26] for cleavage c-terminal of M, 3-Bromo-3-methyl-2-(2-nitrophenylthio)-3H-indole (BNPS-skatole) for cleavage c-terminal of W [27], and 2-nitro-5-thiocyanobenzoic acid (NTCB) for cleavage n-terminal of C [28,29]. Peptide populations were filtered using both length and molecular weight constraints. Since the filtration thresholds affect the proteome coverage prediction, multiple cutoff values are compared. The [R] code is available at: https://github.com/jgmeyerucsd/ProteomeDigestSim.

3. Results and discussion

**Minimum unique peptide length** – The probability of a sequence being unique can be calculated assuming a random distribution of sequences in the library. The number of sequences of length *n* can be described by: $20^n$. Therefore, any given sequence of length five is likely to occur once in a library of 3,200,000 random amino acid sequences (roughly the number of amino acids in the *S. cerevisiae* proteome). As the number of amino acids in the database grows, a peptide sequence must be longer to expect uniqueness. The human proteome contains 11,323,900 amino acids (not including isoforms, downloaded from uniprot on October 22[nd], 2013), and therefore, for a sequence to be unique, it must be length six. Of course, due to common sequence motifs there are less unique peptide sequences in a proteome than would be found in a random library.

**Peptide length distributions from various cleavages** - Initial *in silico* digestions using single cleavage agents were used to compare the resulting peptide lengths (figure 2). Many peptide sequences are too short to uniquely match a protein. For all digestion agents, the most frequent peptide length produced is one. Generation of a single amino acid would arise when the target residue is next to itself in the protein.  Notably, over 25% of theoretical peptides from trypsin digestion, which cleaves after 11.7% of all residues, are of length one.  Not surprisingly, the observable proportion of the residue targeted for cleavage correlates with the resulting average peptide length (figure 3); more common cleavage targets produce shorter average peptide lengths.  Additionally, the residue-level coverage was found to depend on digestion.  Proteome cleavage after more common residues results in depletion of the target residues (figure 4), which is expected to result from production of peptides that are too short to uniquely match a protein sequence.  However, cleavage after rare residues results in enriched coverage of the target residue.  This result was also observed by amino acid analysis of proteome digestions in recent work [30].

**Comparison of peptide filtration parameters –** The theoretical distribution of peptides passing through a MWCO ultrafilter certainly does not match the actual distribution.  Denatured peptides and proteins are effectively larger than folded proteins, and in fact, it was found that even 30kDa or 50kDa cut-off ultrafilters perform better for peptide yield than 10 kDa cut-off ultrafilters [31], despite the inability to identify such large peptide sequences by bottom-up proteomics.  Therefore, multiple length constraints were compared for their influence on the predicted proteome coverage.  Figure 5 shows how various minimum peptide length values affect residue-level depletion and theoretical proteome coverage.  As the minimum length increases, total coverage decreases and depletion of R/K increases.  Figure 6 shows how different upper length thresholds change theoretical coverage.  Intuitively, raising the upper length limit of identifiable peptides increases total predicted proteome coverage.  Interestingly, although total predicted coverage increases, the coverage of R/K stays around 60%.  Since peptide MW also determines identifiable peptides, and peptides above 5kDa are unlikely to be identified with current MSMS technology, an upper limit of 5 kDa was used for subsequent digest simulations. A lower length limit of 7 amino acids was used because this length is more likely to be relevant to actual proteomics experiments.

**Comparison of digestion iterations –** Digest simulations for various digestion iterations were performed to compute theoretical proteome coverage for various iterative digestions.  Simulations confirm that iterative digestion offers theoretically greater coverage of the proteome when the sequence of digestions starts with the protease targeting the rarest residue first (table 1).  As expected, reversal of the optimal digestion sequence results in a negligible improvement to proteome coverage as compared to the limit from using trypsin digestion alone.

**Proposed iterative digestion strategy -** An ideal iterative cleavage strategy must limit sample processing steps, and must take place under conditions that are compatible with the ultrafiltration device.  Further, because tryptophan fluorescence can be used to quantify peptide yield from each digestion, chemical cleavage after tryptophan should initially be omitted since it destroys the fluorophore.  Therefore, an ultrafilter-compatible strategy, with a balance between sample processing and predicted gains in coverage, is the sequence: NTCB, CNBr, LysC, and Trypsin.  Implementation of this method will likely require optimization at various steps.

4. **Conclusions**

This work provides a publically accessible computational framework for simulation of iterative proteome digestion that can be used with any input protein sequence database to optimize proteome coverage. Further, this works demonstrates how the choice of proteome digestion agent affects the predicted proteome coverage due to the distribution of peptide lengths that are produced. This work also shows how various digestion agents affect proteome coverage at the residue level. Proteome cleavage targeting common residues results in depletion of the cleaved residue, but proteome cleavage after rare residues results in enrichment of the target residue. Finally, this paper finds that the best theoretical proteome coverage is achieved by an iterative digestion strategy that limits production of short peptides by cleaving the rarest residues first.

**Conflicts of interest**

The author declares that there is no conflict of interests regarding the publication of this article.

**Acknowledgements**

JGM was supported by the NIH interfaces training grant (T32EB009380) and funding from the NSF (MCB1244506).

**References**

[1] Y. Zhang, B. R. Fonslow, B. Shan, M.-C. Baek, and J. R. Yates, "Protein Analysis by Shotgun/Bottom-up Proteomics," *Chem. Rev.*, vol. 113, no. 4, pp. 2343–2394, Feb. 2013.
[2] B. Meyer, D. G. Papasotiriou, and M. Karas, "100% protein sequence coverage: a modern form of surrealism in proteomics," *Amino Acids*, vol. 41, no. 2, pp. 291–310, Jul. 2010.
[3] J. M. Gilmore and M. P. Washburn, "Advances in shotgun proteomics and the analysis of membrane proteomes," *Journal of Proteomics*, vol. 73, no. 11, pp. 2078–2091, Oct. 2010.

[4] M. Rey, H. Mrázek, P. Pompach, P. Novák, L. Pelosi, G. Brandolin, E. Forest, V. Havlíček, and P. Man, "Effective Removal of Nonionic Detergents in Protein Mass Spectrometry, Hydrogen/Deuterium Exchange, and Proteomics," *Anal. Chem.*, vol. 82, no. 12, pp. 5107–5116, May 2010.

[5] A. Motoyama and J. R. Yates, "Multidimensional LC Separations in Shotgun Proteomics," *Anal. Chem.*, vol. 80, no. 19, pp. 7187–7193, 2008.

[6] Y. Wang, F. Yang, M. A. Gritsenko, Y. Wang, T. Clauss, T. Liu, Y. Shen, M. E. Monroe, D. Lopez-Ferrer, T. Reno, R. J. Moore, R. L. Klemke, D. G. Camp, and R. D. Smith, "Reversed-phase chromatography with multiple fraction concatenation strategy for proteome profiling of human MCF10A cells," *Proteomics*, vol. 11, no. 10, pp. 2019–2026, 2011.

[7] L. H. Betancourt, P.-J. De Bock, A. Staes, E. Timmerman, Y. Perez-Riverol, A. Sanchez, V. Besada, L. J. Gonzalez, J. Vandekerckhove, and K. Gevaert, "SCX charge state selective separation of tryptic peptides combined with 2D-RP-HPLC allows for detailed proteome mapping," *Journal of Proteomics*, vol. 91, no. 0, pp. 164–171, Oct. 2013.

[8] A. Michalski, E. Damoc, J.-P. Hauschild, O. Lange, A. Wieghaus, A. Makarov, N. Nagaraj, J. Cox, M. Mann, and S. Horning, "Mass Spectrometry-based Proteomics Using Q Exactive, a High-performance Benchtop Quadrupole Orbitrap Mass Spectrometer," *Molecular & Cellular Proteomics*, vol. 10, no. 9, Sep. 2011.

[9] J. V. Olsen, J. C. Schwartz, J. Griep-Raming, M. L. Nielsen, E. Damoc, E. Denisov, O. Lange, P. Remes, D. Taylor, M. Splendore, E. R. Wouters, M. Senko, A. Makarov, M. Mann, and S. Horning, "A Dual Pressure Linear Ion Trap Orbitrap Instrument with Very High Sequencing Speed," *Molecular & Cellular Proteomics*, vol. 8, no. 12, pp. 2759–2769, Dec. 2009.

[10] C. K. Frese, A. F. M. Altelaar, M. L. Hennrich, D. Nolting, M. Zeller, J. Griep-Raming, A. J. R. Heck, and S. Mohammed, "Improved Peptide Identification by Targeted Fragmentation Using CID, HCD and ETD on an LTQ-Orbitrap Velos," *J. Proteome Res.*, vol. 10, no. 5, pp. 2377–2388, 2011.

[11] R. J. Chalkley, P. R. Baker, K. F. Medzihradszky, A. J. Lynn, and A. L. Burlingame, "In-depth Analysis of Tandem Mass Spectrometry Data from Disparate Instrument Types," *Molecular & Cellular Proteomics*, vol. 7, no. 12, pp. 2386–2398, Dec. 2008.

[12] Y. Shen, N. Tolić, S. O. Purvine, and R. D. Smith, "Improving Collision Induced Dissociation (CID), High Energy Collision Dissociation (HCD), and Electron Transfer Dissociation (ETD) Fourier Transform MS/MS Degradome–Peptidome Identifications Using High Accuracy Mass Information," *J. Proteome Res.*, vol. 11, no. 2, pp. 668–677, Nov. 2011.

[13] S. Kim, N. Mischerikow, N. Bandeira, J. D. Navarro, L. Wich, S. Mohammed, A. J. R. Heck, and P. A. Pevzner, "The generating function of CID, ETD and CID/ETD pairs of tandem mass spectra: Applications to database search," *Molecular & Cellular Proteomics*, 2010.

[14] S. Linnarsson, "Recent advances in DNA sequencing methods – general principles of sample preparation," *Experimental Cell Research*, vol. 316, no. 8, pp. 1339–1343, May 2010.

[15] B. Rietschel, T. N. Arrey, B. Meyer, S. Bornemann, M. Schuerken, M. Karas, and A. Poetsch, "Elastase Digests," *Molecular & Cellular Proteomics*, vol. 8, no. 5, pp. 1029–1043, May 2009.

[16] G. Choudhary, S.-L. Wu, P. Shieh, and W. S. Hancock, "Multiple Enzymatic Digestion for Enhanced Sequence Coverage of Proteins in Complex Proteomic Mixtures Using Capillary LC with Ion Trap MS/MS," *J. Proteome Res.*, vol. 2, no. 1, pp. 59–67, Nov. 2002.

[17] H. Moura, R. R. Terilli, A. R. Woolfitt, Y. M. Williamson, G. Wagner, T. A. Blake, M. I. Solano, and J. R. Barr, "Proteomic Analysis and Label-Free Quantification of the Large Clostridium difficile Toxins," *International Journal of Proteomics*, vol. 2013, pp. 1–10, 2013.

[18] B. Q. Tran, C. Hernandez, P. Waridel, A. Potts, J. Barblan, F. Lisacek, and M. Quadroni, "Addressing Trypsin Bias in Large Scale (Phospho)proteome Analysis by Size Exclusion Chromatography and Secondary Digestion of Large Post-Trypsin Peptides," *J. Proteome Res.*, vol. 10, no. 2, pp. 800–811, Dec. 2010.

[19] N. Neuhauser, N. Nagaraj, P. McHardy, S. Zanivan, R. Scheltema, J. Cox, and M. Mann, "High Performance Computational Analysis of Large-scale Proteome Data Sets to Assess Incremental Contribution to Coverage of the Human Genome," *J. Proteome Res.*, vol. 12, no. 6, pp. 2858–2868, Apr. 2013.

[20] D. L. Swaney, C. D. Wenger, and J. J. Coon, "Value of Using Multiple Proteases for Large-Scale Mass Spectrometry-Based Proteomics," *Journal of Proteome Research*, vol. 9, no. 3, pp. 1323–1329, Mar. 2010.

[21] J. R. Wiśniewski and M. Mann, "Consecutive Proteolytic Digestion in an Enzyme Reactor Increases Depth of Proteomic and Phosphoproteomic Analysis," *Anal. Chem.*, vol. 84, no. 6, pp. 2631–2637, Feb. 2012.

[22] J. R. Wisniewski, A. Zougman, N. Nagaraj, and M. Mann, "Universal sample preparation method for proteome analysis," *Nat Meth*, vol. 6, no. 5, pp. 359–362, May 2009.

[23] R Development Core Team, *R: A language and environment for statistical computing*. Vienna, Austria: R Foundation for Statistical Computing, 2008.

[24] D. L. Crimmins, S. M. Mische, and N. D. Denslow, "Chemical Cleavage of Proteins in Solution," in *Current Protocols in Protein Science*, John Wiley & Sons, Inc., 2001.

[25] R. Kaiser and L. Metzka, "Enhancement of Cyanogen Bromide Cleavage Yields for Methionyl-Serine and Methionyl-Threonine Peptide Bonds," *Analytical Biochemistry*, vol. 266, no. 1, pp. 1–8, Jan. 1999.

[26] Y. A. Andreev, S. A. Kozlov, A. A. Vassilevski, and E. V. Grishin, "Cyanogen bromide cleavage of proteins in salt and buffer solutions," *Analytical Biochemistry*, vol. 407, no. 1, pp. 144–146, Dec. 2010.

[27] M. M. Vestling, M. A. Kelly, C. Fenselau, and C. E. Costello, "Optimization by mass spectrometry of a tryptophan-specific protein cleavage reaction," *Rapid Commun. Mass Spectrom.*, vol. 8, no. 9, pp. 786–790, Sep. 1994.

[28] G. R. Jacobson, M. H. Schaffer, G. R. Stark, and T. C. Vanaman, "Specific Chemical Cleavage in High Yield at the Amino Peptide Bonds of Cysteine and Cystine Residues," *Journal of Biological Chemistry*, vol. 248, no. 19, pp. 6583–6591, Oct. 1973.

[29] M. Iwasaki, T. Masuda, M. Tomita, and Y. Ishihama, "Chemical Cleavage-Assisted Tryptic Digestion for Membrane Proteome Analysis," *J. Proteome Res.*, vol. 8, no. 6, pp. 3169–3175, Jun. 2009.

[30] J. G. Meyer, S. Kim, D. Maltby, M. Ghassemian, N. Bandeira, and E. A. Komives, "Expanding proteome coverage with orthogonal-specificity alpha-lytic proteases," *Molecular & Cellular Proteomics*, Jan. 2014.

[31] J. R. Wisniewski, D. F. Zielinska, and M. Mann, "Comparison of ultrafiltration units for proteomic and N-glycoproteomic analysis by the filter-aided sample preparation method," *Analytical Biochemistry*, vol. 410, no. 2, pp. 307–309, Mar. 2011.

**Figure 1: Cartoon describing the multiple-enzyme digestion, filter-assisted sample preparation strategy (MED-FASP) from Wiesinski and Mann.** A proteome is digested on top of a size-based filter device and peptides are then spun through the filter. Undigested sequences are retained above the filter because of their length. The process is repeated with various cleavage agents and several peptide pools are collected separately. The peptides are then analyzed by nLC-MS/MS separately and the resulting data is then combined either before or after the database search.

**Figure 2: Theoretical peptide length distributions produced from various cleavage agents.** (A) Size frequency distributions (density) of peptides from proteome digestion by five real (i.e. trypsin, LysC, GluC, CNBr, NTCB) and one theoretical cleavage agent (LeuC). The vertical black lines at 7 and 35 indicate general peptide identification size limits. (B) The same distribution focused on the region from 1-10 amino acids. (C) The view focused on the region between 30-40 amino acids.

**Figure 3: Correlation between abundance of the residue targeted for cleavage and the resulting average peptide length.** Proteome cleavage targeting abundant residues result in lower average peptide lengths; proteome cleavage targeting rare residues results in higher average peptide length. The line shows the data fit to an exponential equation.

**Figure 4: Residue-level coverage observed for various cleavage agents.** Proteome cleavage of more common amino acids, such as with (A) trypsin or the theoretical cleavage after (B) Leucine, result in residue-specific depletion of the target residues. However, cleavage of rare amino acids, such as (C) Methionine or (D) Cysteine, results in residue-specific enrichment of the target residues.

**Figure 5: Effect of minimum peptide length on proteome coverage and residue-level depletion.** Residue-level coverage predicted after trypsin digestion keeping all peptides with lengths between: (A) 1-35, (B) 5-35, (C) 7-35, and (D) 10-35.

**Figure 6: Effect of upper length limit on predicted proteome coverage.** Theoretical coverage keeping peptides with length (A) 5-20, (B) 5-30, (C) 5-40, and (D) 5-100 residues. As the upper length limit increases, the theoretical coverage maximum increases.

**Table 1 legend: Theoretical upper limits of coverage upon digestion with various cleavage agents using the MED-FASP strategy.** Iterative cleavage of the proteome starting with the rarest amino acids first results in the greatest theoretical proteome coverage of 91.1%. The reversed sequence of cleavage provides a minimal improvement to theoretical proteome coverage. Peptides were filtered after each digest keeping those with MW >5 kDa for additional digestion.

Figure 1: MED-FASP digestion strategy

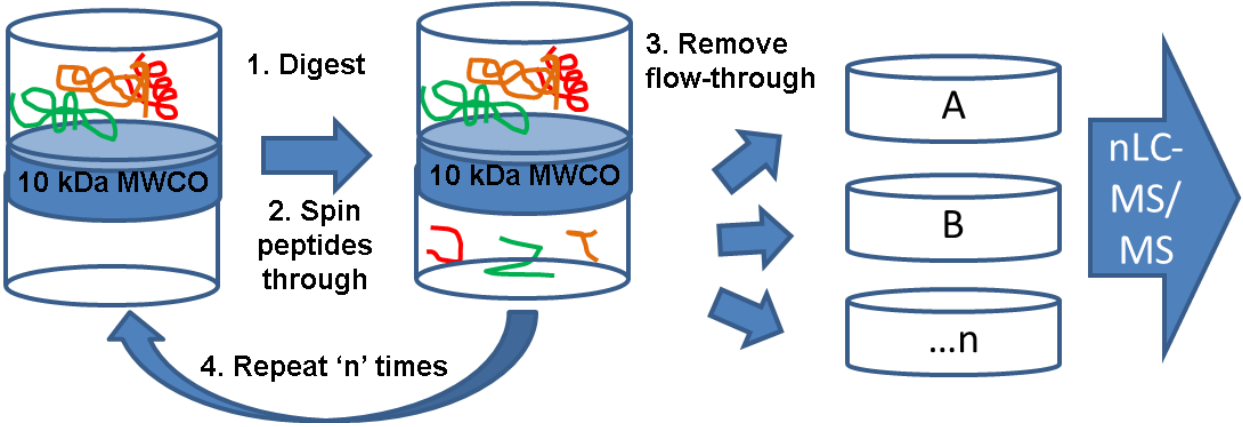

Figure 2: Theoretical peptide lengths upon digestions with various specificities

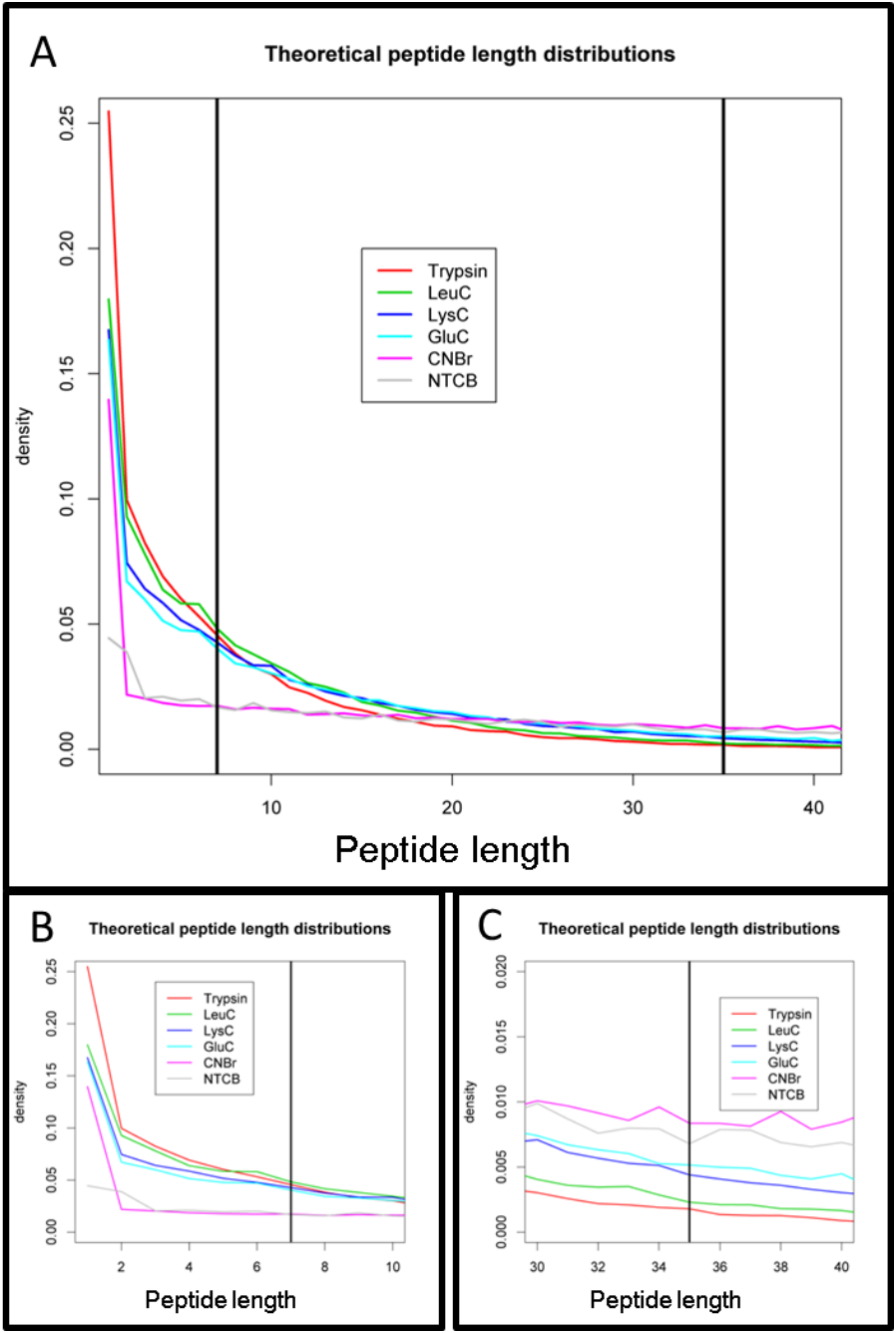

Figure 3: Residue specificity predicts average peptide length

| | % in *S. cerevisiae* proteome |
|---|---|
| **L** | **9.56** |
| S | 9.04 |
| **K** | **7.29** |
| I | 6.57 |
| **E** | **6.45** |
| N | 6.12 |
| T | 5.91 |
| D | 5.77 |
| V | 5.57 |
| A | 5.48 |
| G | 4.95 |
| F | 4.50 |
| **R** | **4.45** |
| P | 4.39 |
| Q | 3.92 |
| Y | 3.39 |
| H | 2.18 |
| **M** | **2.10** |
| **C** | **1.31** |
| W | 1.05 |

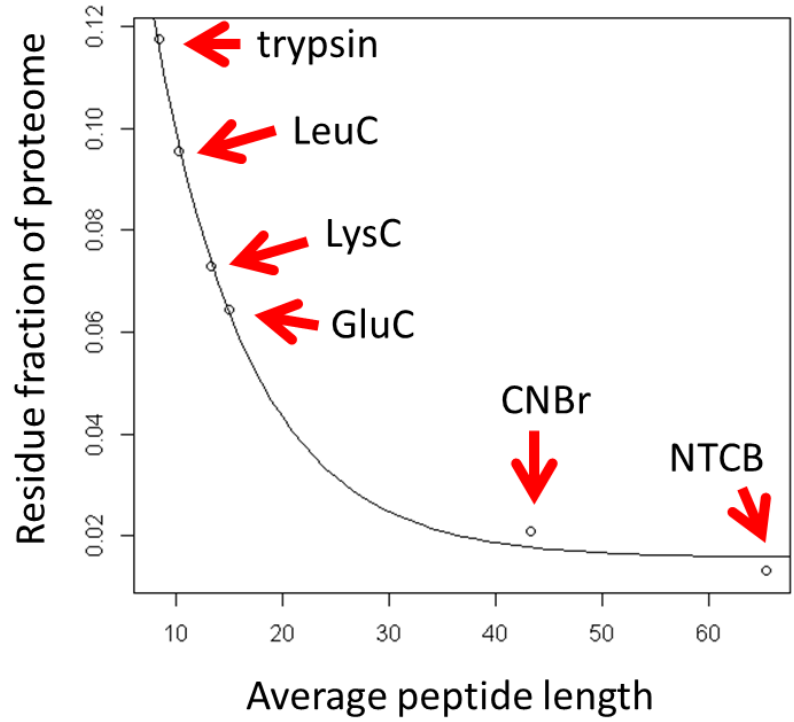

Figure 4: Predicted residue-level proteome coverage from various digestions

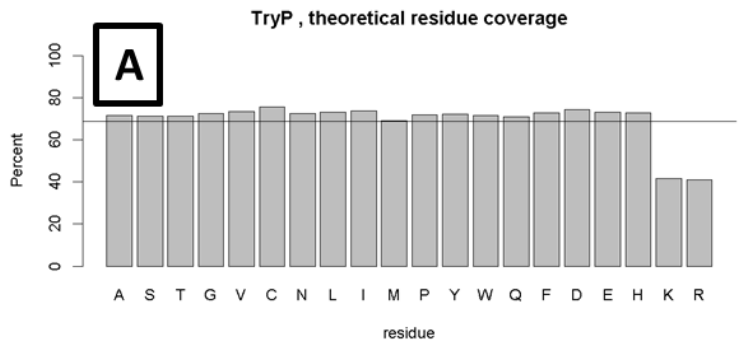

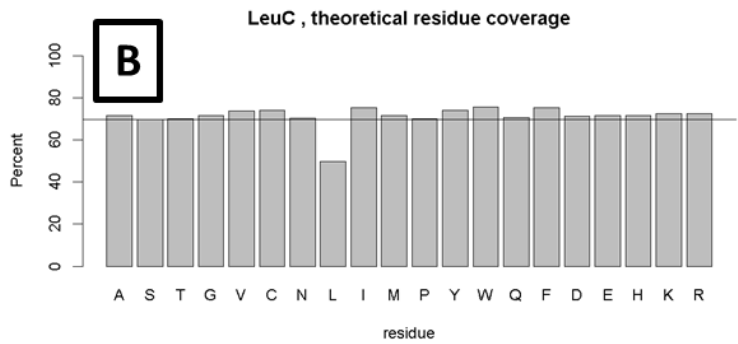

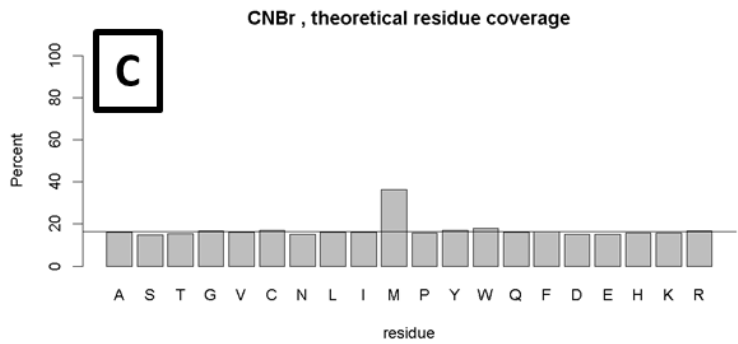

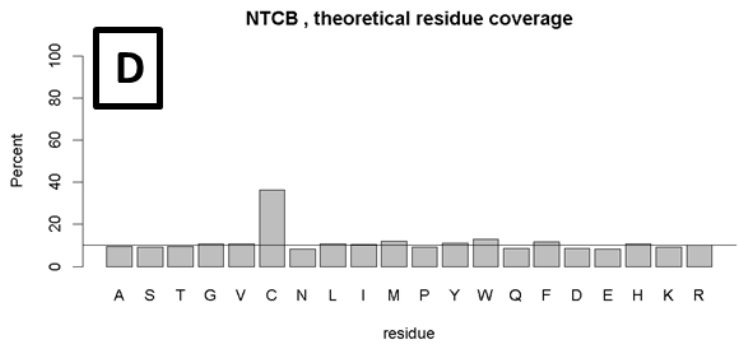

Figure 5: Effect of minimum peptide length on proteome coverage and residue-level depletion

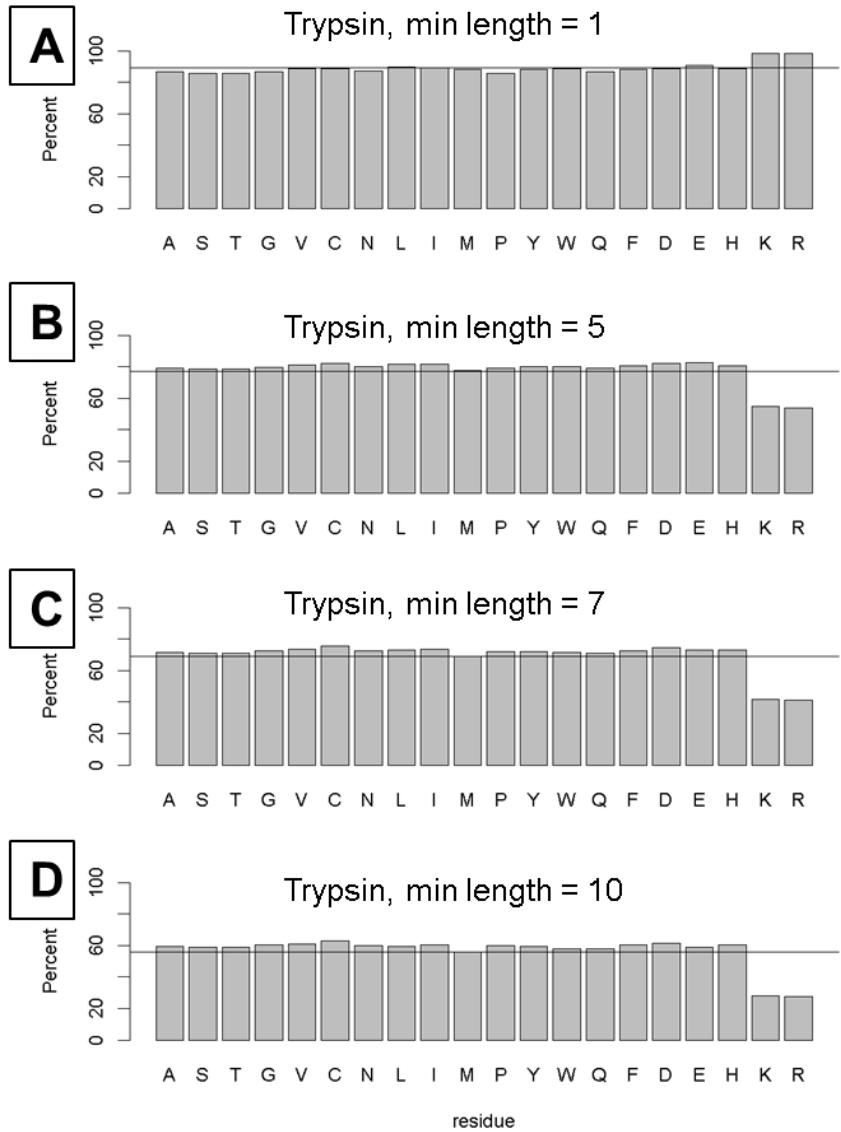

Figure 6

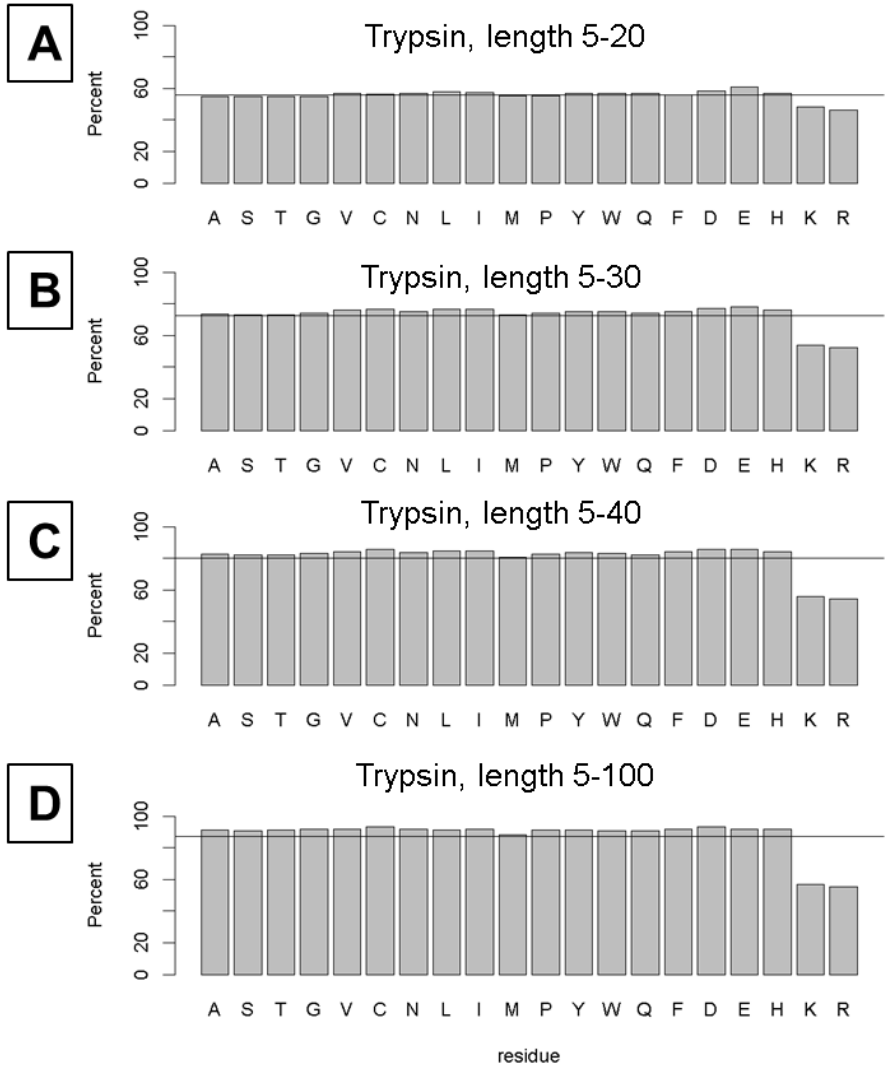

Table 1: Comparison of theoretical coverage resulting from various digestions

| Digestion strategy | Theoretical coverage limit (%) |
| --- | --- |
| Trypsin | 69.5 |
| LysC | 67.1 |
| GluC | 62.7 |
| AspN | 63.1 |
| ArgC | 52.4 |
| CNBr | 22.4 |
| NTCB | 13.6 |
| TrpC | 10.9 |
| LysC, Trypsin | 81.2 |
| GluC, Trypsin | 81.1 |
| CNBr, LysC, Trypsin | 84.4 |
| NTCB, CNBr, LysC, Trypsin | 86.3 |
| TrpC, NTCB, CNBr, ArgC, GluC, Trypsin | 87.9 |
| TrpC, NTCB, CNBr, ArgC, AspN, GluC, Trypsin | 91.1 |
| Trypsin, GluC, AspN, ArgC, CNBr, NTCB, TrpC[a] | 74.2 |

[a]reversed order relative to optimal